\documentclass[a4paper,9pt]{article}

\newcommand{\brabra}{{\langle\!\langle}}
\newcommand{\ketket}{{\rangle\!\rangle}}

\newcommand{\A}{{\mathcal A}}

\makeatletter
\@addtoreset{equation}{section}

\makeatother

\usepackage{multirow}
\usepackage{booktabs}
\usepackage{graphicx}
\usepackage{amsmath}
\usepackage{amsthm}
\usepackage{amssymb}
\usepackage{eepic}
\usepackage{amscd}
\usepackage{longtable}
\usepackage{array}

\newcommand{\nc}{\newcommand}
\nc{\rnc}{\renewcommand}
\nc{\nn}{\nonumber}

\newcommand{\del}{\delta}
\newcommand{\eps}{\epsilon}
\newcommand{\one}{\mbox{1 \hspace{-3.2mm} {\bf \rm l}} }

\newcommand{\ba}{{{\mathbf a}}}

\newcommand{\ot}{{\otimes  }}

\newcommand{\bsq}{\blacksquare}
\newcommand{\sq}{\square}

\begin{document}

\title{Matrix product solution to   
 an inhomogeneous multi-species  TASEP }

\author{
 Chikashi Arita and Kirone Mallick\\ 
 {\normalsize Institut de Physique Th\'{e}orique, CEA Saclay,
 F-91191 Gif-sur-Yvette, France }
}
 
\date{\today}

\maketitle

\begin{abstract}
 We study a multi-species exclusion process with inhomogeneous hopping rates.
This model is equivalent to a  Markov chain on the symmetric group that corresponds to a random walk in the affine braid arrangement \cite{Lam}.  We find a matrix product representation for the stationary state of this  model.
We also show that it is equivalent to a graphical construction proposed by Ayyer and Linusson
\cite{AL}, which generalizes Ferrari and  Martin's construction \cite{FM}.
\end{abstract}

\section{Introduction}\label{sec:intro}

The coupling  of  randomness with   algebraic or arithmetic
 structures can lead to beautiful combinatorial results.
 The study of random
 partitions of integers and its extension to three-dimensional
 plane partitions  \cite{Vershik,Andrews,Okounkov} provides a 
 prominent  example. Recently,  Lam studied  Markov chains
 that can be represented geometrically  as  random walks on  a 
 regular tessellation of a  vector space on which an  affine 
 Weyl  group  acts \cite{Lam}.
When this  Weyl  group  is a symmetric group, the stationary distribution of the chain displays remarkable combinatorial properties that were further  explored by Lam and Williams, leading to various conjectures \cite{LW}.
The Markov chain studied in \cite{LW} is equivalent
 to a multi-species exclusion process with inhomogeneous transition rates
 as shown by Ayyer and Linusson in a very recent publication \cite{AL}.
 This suggests  that 
the powerful techniques that were developed in non-equilibrium
 statistical mechanics to analyze the asymmetric exclusion process 
 \cite{Li99, DEHP, BE} should be relevant to this more mathematical problem.
 Indeed,  Ayyer and   Linusson   conjectured that 
the stationary state can be obtained by generalizing the elegant
 graphical  algorithm, invented by 
Ferrari and Martin  \cite{FM} to solve the homogeneous $N$-species 
 totally  asymmetric simple exclusion process ($N$-TASEP); 
 in \cite{AL},  they  generalized Ferrari and Martin's algorithm
 and  explored the consequences
 for  Lam and Williams' conjectures.

 In the present work, we solve the  $N$-TASEP
 with inhomogeneous transition rates by using a generalized matrix product Ansatz.
 More precisely,  we show that a suitable deformation of the homogeneous algebra studied in \cite{PEM,AAMP1} allows us to calculate the stationary state of the inhomogeneous $N$-TASEP. 
 We also show that our matrix product solution is  equivalent to
 the algorithm conjectured  in  \cite{AL}.

 The outline of this work  is as follows. In section \ref{sec:def},
 the model is defined. In section \ref{sec:MPA},
 we give a matrix product solution to the stationary state. In 
  section \ref{sec:FMALalgo}, we explain the graphical construction
 of the stationary state conjectured by Ayyer and Linusson,
 and we show that this construction is equivalent to the matrix product solution.
 We give concluding remarks in  section \ref{sec:conc}.

\section{Definition of the model}\label{sec:def}

 We consider an $L$-site  periodic chain in which 
each site takes a non-negative integer value
$1,2,\dots,$ or $N+1$ (this is called the {\it N-species} problem).
Each pair of nearest neighbor sites 
exchanges their values according to the following
 continuous time  stochastic dynamics:
\begin{eqnarray}
  a\ b \ \to b\ a \quad  {\rm with\ rate}
  \left\{\begin{array}{ll} x_a & (a<b) \\ 0 & ({\rm otherwise})
   \end{array}\right.
\end{eqnarray}
The special case  $x_a=1$ for all $a$ corresponds to the 
{\it homogeneous} $N$-TASEP \cite{AAMP1,AKSS,FM,EFM}.
 When the transition rates  $x_a$'s are not equal to each other, we say
 that this dynamics is {\it inhomogeneous}  
 (although one could  consider even 
more complicated or general transition rules).
When $L=N+1$, and there is exactly one particle of each species
in the system, the process is equivalent to Lam and Williams' process on
the symmetric group $S_L$ \cite{LW}.

The  inhomogeneous  $N$-TASEP is governed by 
the master equation
\begin{eqnarray}
 \frac{d}{dt}|P\rangle =M^{(N)}|P\rangle 
\end{eqnarray}
for the vector $|P\rangle=\sum P(j_1\cdots j_L)|j_1\cdots j_L\rangle$,
where $P(j_1\cdots j_L)$ is the probability of finding the system in 
 a configuration $j_1\cdots j_L$.
The generator matrix (Markov matrix) $M^{(N)}$  is the summation 
 of local operators $\left( M^{(N)}_{\rm Loc} \right)_{i,i+1}$ that acts on 
the spaces  corresponding to  $i$th and $(i+1)$st  sites of the chain:
\begin{eqnarray}
  M^{(N)} = \sum_{i=1}^L \left( M^{(N)}_{\rm Loc} \right)_{i,i+1},\ 
   M^{(N)}_{\rm Loc}  =
\sum_{  a,b=1}^{  N+1}  \Theta(a-b) \Big( | ba \rangle\langle ab  |
 -    | ab \rangle\langle ab  | \Big) \  \hbox{ with }
 \Theta (a-b) = \left\{\begin{array}{ll}
  x_a & (a<b), \\  0 & (a\ge b) .  \end{array}\right.
\end{eqnarray}

We write $m=(m_1,\dots,m_{N+1}) \ (\hbox{with } m_i\in\mathbb Z_{\ge 0})$
 for the sector that  contains $m_k$ particles of type $k$.
Because of the  conservation of the number of particles of each type,
 we have the decomposition $ M^{(N)} =  \bigoplus_m M_m$.
In particular,  we consider {\it basic sectors}, i.e. $m_i>0$.   
Since the identification of local states
$N+1\to N$ maps the $N$-species dynamics 
to the $(N-1)$-species dynamics,
we have the spectral inclusion:
the spectrum of the sector 
$m=(m_1,\dots,m_N,m_{N+1} )$
contains that of the sector
$m'= (m_1,\dots,m_N+m_{N+1} )$.\footnote{
More general inclusion relations are satisfied 
for the homogeneous case, see \cite{AKSS}.}
This inclusion relation indicates that 
there exists a ``conjugation matrix'' $\psi_m$ such that 
\begin{eqnarray}\label{eq:conjugation-relation}
   M_m \psi_m = \psi_m M_{m'}  .
\end{eqnarray} 
 This matrix  allows one to lift states
 that belong to the $(N-1)$-species sector $m'$ and to construct
 eigenstates for the  $N$-species sector $m$. In particular,
 the stationary state of the $N$-TASEP (the kernel of $M^{(N)}$)
 can be constructed recursively if one knows the sequence
 of the conjugation matrices.

\section{Generalized matrix product Ansatz} \label{sec:MPA}

The idea of the  matrix product Ansatz is to express 
the probability of finding each configuration $j_1\cdots j_L$
 as a trace over a suitable algebra
\begin{eqnarray}\label{eq:MP}
  P(j_1\cdots j_L) = \frac{1}{Z}{\rm Tr }  \Big(X^{(N)}_{j_1}\cdots X^{(N)}_{j_L} \Big) 
\end{eqnarray}
with    a normalization constant $Z$.
The operators $X^{(N)}_{j}$'s
 must satisfy suitable algebraic relations,  which are usually infinite dimensional 
  \cite{DEHP,BE}.
For the homogeneous $N$-TASEP, the  operators  $X^{(N)}_{j}$ were constructed recursively
\cite{EFM,PEM} as
tensor products of four fundamental operators  $\del,\eps,A$ and $\one$ 
 that act on an infinite dimensional space
 $\A = \bigotimes_{\mu\ge 0} {\mathbb C} |\mu\ketket $ as 
\begin{eqnarray}
    \del|\mu\ketket=\left\{ \begin{array}{ll}
  0 & (\mu=0) , \\ |\mu-1\ketket & (\mu>0)  ,
 \end{array}\right.\    
    \eps|\mu\ketket=|\mu+1\ketket,\  
 A|\mu\ketket=
\left\{ \begin{array}{ll}
  |0 \ketket & (\mu=0) , \\   0 & (\mu>0)  ,
 \end{array}\right.\  
 \one |\mu\ketket =|\mu\ketket  .
\label{eq:actions}
\end{eqnarray}
One can easily verify that the following quadratic relations are satisfied:
\begin{eqnarray}
  \del \eps = \one,\ \del A =0,\ A\eps =0 \, .
\label{eq:DEHP}
\end{eqnarray}
 In \cite{PEM},  the stationary state 
 for the homogeneous 
 $N$-TASEP has been expressed as a matrix product form, using an algebraic
 interpretation of Ferrari and Martin's algorithm \cite{FM}. 
 More recently   the matrix  product Ansatz technique
 was  generalized  to obtain a conjugation matrix 
 that  satisfies  ``conjugation relation''
(\ref{eq:conjugation-relation}) between systems  having different numbers of species
 \cite{AAMP1}.
 We now explain how the same ideas can be adapted   to the  inhomogeneous case.

 The stationary state for $N=1$ case is trivial,
i.e. all the possible states in each sector 
are realized with a same probability.
This can be regarded as one dimensional representation
of the matrices $X_1^{(1)}=X_2^{(1)}=1$.

 \subsection{The 2 species  case}

 For $N=2$, a
 matrix product stationary representation has been known, even in the inhomogeneous
 case \cite{BE}. One  possible choice for the algebra is
\begin{eqnarray}
   X^{(2)}_1X^{(2)}_3 = \frac{1}{x_1}X^{(2)}_1 + X^{(2)}_3, \quad
   X^{(2)}_2X^{(2)}_3 = \frac{1}{x_2}X^{(2)}_2,\quad 
   X^{(2)}_1X^{(2)}_2=X^{(2)}_2. 
\end{eqnarray} 
Defining an operator-valued vector
$ \mathbf X^{(2)}  =
 \left(\begin{array}{c}  X^{(2)}_1 \\ X^{(2)}_2 \\ X^{(2)}_3 \end{array}\right)   $,
we  observe that the following  decomposition exists
\begin{eqnarray}
 \mathbf X^{(2)}  =  \left(\begin{array}{cc}
   \one & \del \\ 0 & A \\ y_1 \eps &  y_1B + y_2A \end{array}\right)  
  \left(\begin{array}{c}  1  \\ 1  \end{array}\right) \,,
\label{DecompX2}
\end{eqnarray}
where we have set $y_1 = 1/x_1, y_2= 1/x_2$ and the matrices
$\del, \eps, A$ satisfy~(\ref{eq:DEHP}). This decomposition
 can be written more formally as the product of two rectangular operator-valued
 matrices  $\ba^{(1)}$  and   $\ba^{(2)}$  of sizes $2 \times 1$ and $3 \times 2$ respectively
\begin{eqnarray}
 \mathbf X^{(2)}   =  \left(\begin{array}{cc}
   a_{11}^{(2)} & a_{12}^{(2)} \\
    a_{21}^{(2)} & a_{22}^{(2)} \\ a_{31}^{(2)} &  a_{32}^{(2)} 
    \end{array}\right)  
  \left(\begin{array}{c}  a_{11}^{(1)} \\ a_{21}^{(1)}  \end{array}\right)  
  := \ba^{(2)} \star   \mathbf a^{(1)} \, , 
\label{FORMALDecompX2}
\end{eqnarray}
 where the symbol $\star$ means that we perform tensor products amongst the elements
 of the matrices  $\ba^{(2)}$   and  $\ba^{(1)}$.
 More generally, for  matrix-valued matrices $Y=(Y_{uv})_{uv},Z =(Z_{uv})_{uv}$, 
it represents the  product 
$Y\star Z=\left(\sum_{w}  Y_{uw}\ot Z_{wv}\right)_{uv}$.
Since   $a_{11}^{(1)}$  and $a_{21}^{(1)}$
 are  scalars here,
 the tensor product reduces to the ordinary product.
 Finally, we observe that the  matrices $\ba^{(1)}$
 and $\ba^{(2)}$ satisfies 
\begin{eqnarray}
\label{Intertw1}
  M^{(1)}_{\rm Loc}\ba^{(1)} \ot \ba^{(1)} - \ba^{(1)}\ot\ba^{(1)} M^{(0)}_{\rm Loc}
 =\widehat \ba^{(1)}\ot \ba^{(1)} - \ba^{(1)} \ot \widehat \ba^{(1)}  \\
  M^{(2)}_{\rm Loc}\ba^{(2)} \ot \ba^{(2)} - \ba^{(2)}\ot\ba^{(2)} M^{(1)}_{\rm Loc}
 =\widehat \ba^{(2)}\ot \ba^{(2)} - \ba^{(2)} \ot \widehat \ba^{(2)}   
\label{Intertw2}
\end{eqnarray}
where 
\begin{eqnarray}
 \widehat \ba^{(1)}   =
   \left(\begin{array}{c}  0  \\ x_1  \end{array}\right) ,\quad 
 \widehat \ba^{(2)}   =
   \left(\begin{array}{cc}  0 & 0  \\  0 & 0 \\ \eps & \one \end{array}\right) .
\end{eqnarray}
Since $M^{(0)}_{\rm Loc}=0$,   equation (\ref{Intertw1})
 is the relation in the usual matrix product Ansatz \cite{BE}.
 The relation  (\ref{Intertw2})  implies that the  matrix $\psi_m$ whose 
elements are given as 
\begin{eqnarray}
 \langle j_1\cdots j_L  |  \psi_m  | k_1\cdots k_L \rangle
   = {\rm Tr} \Big(a^{(2)}_{j_1k_1}\cdots a^{(2)}_{j_Lk_L} \Big)
\end{eqnarray}
intertwines the dynamics 
of the sectors $m=(m_1,m_2,m_3)$ and $m'=(m_1,m_2+m_3)$,  i.e. 
it satisfies equation  (\ref{eq:conjugation-relation}).
Formally we write  the stationary state for the sector $m'$
as $\psi_{ m' }$ with $ \langle j_1\cdots j_L  |  \psi_{m'}  | 1\cdots 1 \rangle
   = {\rm Tr} \Big(a^{(1)}_{j_11}\cdots a^{(1)}_{j_L1} \Big) =1 $.
Then  the stationary state for the sector $m$ can be rewritten as 
$\psi_m \psi_{ m' } $.

For a simple nontrivial example, 
the stationary state of the sector $(1,1,2)$ is given as 
\begin{eqnarray}
\psi_{ (1,1,2) } \psi_{ (1,3) }  = 
  \bordermatrix{
\ & {\scriptstyle 1222} & {\scriptstyle 2122}  &
 {\scriptstyle 2212}  & {\scriptstyle 2221}  \cr 
{\scriptstyle 1233} & y_2^2 & 0 & 0 & 0 \cr
{\scriptstyle 1323} & y_2^2 & y_1 y_2 & 0 & 0 \cr
{\scriptstyle 1332} & y_2^2 & y_1 y_2 & y_1^2 & 0 \cr
{\scriptstyle 2133} & 0 & y_2^2 & y_1 y_2 & y_1^2 \cr
{\scriptstyle 2313} & 0 & 0 & y_2^2 & y_1 y_2 \cr
{\scriptstyle 2331} & 0 & 0 & 0 & y_2^2 \cr
{\scriptstyle 3123} & 0 & y_2^2 & 0 & 0 \cr
{\scriptstyle 3132} & 0 & y_2^2 & y_1 y_2 & 0 \cr
{\scriptstyle 3213} & y_1^2 & 0 & y_2^2 & y_1 y_2 \cr
{\scriptstyle 3231} & y_1 y_2 & 0 & 0 & y_2^2 \cr
{\scriptstyle 3312} & 0 & 0 & y_2^2 & 0 \cr
{\scriptstyle 3321} & y_1 y_2 & y_1^2 & 0 & y_2^2 }
\left(\begin{array}{c}     1 \\  1 \\ 1 \\ 1  \end{array}\right)
=\left(\begin{array}{c}
 y_2^2 \\
 y_2 \left(y_1+y_2\right) \\
 y_1^2+y_1 y_2+y_2^2 \\
 y_1^2+y_1 y_2+y_2^2 \\
 y_2 \left(y_1+y_2\right) \\
 y_2^2 \\
 y_2^2 \\
 y_2 \left(y_1+y_2\right) \\
 y_1^2+y_1 y_2+y_2^2 \\
 y_2 \left(y_1+y_2\right) \\
 y_2^2 \\
 y_1^2+y_1 y_2+y_2^2 \\
\end{array}\right) .
\end{eqnarray}

\subsection{The $N$-species  case}

We now explain the generalized matrix product Ansatz
 for  general values of $N$   (see \cite{AAMP1} for  more details).
Reversing the construction in the last subsection,
 we start  with the  following relation  that  we call ``hat relation'':
\begin{eqnarray}\label{eq:hat}
  M^{(N)}_{\rm Loc}\ba^{(N)} \ot \ba^{(N)}
  -\ba^{(N)}\ot\ba^{(N)} M^{(N-1)}_{\rm Loc}
 = \widehat\ba^{(N)}\ot \ba^{(N)} -
   \ba^{(N)}\ot  \widehat \ba^{(N)} \, ,
\label{N-HATrelation}
\end{eqnarray}
where $\ba^{(N)}$ and  $\widehat \ba^{(N)}$
are operator-valued matrices of size $(N+1)\times N$. 
We write their elements as 
$\langle j|\ba^{(N)}|k\rangle=a^{(N)}_{jk} $, 
$\langle j|\widehat\ba^{(N)}|k\rangle=\widehat a^{(N)}_{jk}$.
We know that, if we can  construct a couple
 $\{\ba^{(N)},\widehat\ba^{(N)}\}$
 (for the general integer of $N$)
that satisfies \eqref{eq:hat},
the matrix $\psi_m$ defined as 
\begin{eqnarray}\label{eq:psi-element}
 \langle j_1\cdots j_L |  \psi_m  | k_1\cdots k_L\rangle
 ={\rm Tr} \Big( a^{(N)}_{j_1k_1} \cdots a^{(N)}_{j_Lk_L} \Big)
\end{eqnarray}
satisfies the conjugation relation (\ref{eq:conjugation-relation}).
Here the  configurations $j_1\cdots j_L $ and $ k_1\cdots k_L$
belong  to the sectors $m=(m_1,\dots,m_{N+1})$
and $m' =(m_1,\dots,m_N+ m_{N+1})$,
respectively. Furthermore, the  stationary state of the sector 
$m$ can be written by the product of 
conjugation matrices
\begin{eqnarray}
\label{eq:psipsipsi}
|\bar P\rangle_m =
\psi_m \psi_{m'} \cdots \psi_{ (m_1,L-m_1) } 
\end{eqnarray} 
if all of them are nonzero
(we note that the  conjugation matrix  $\psi_m $ lifts up
other eigenstates from lower sectors as well).

We set $y_i=1/x_i$ and define the operator  $B=\one-A$.   
 An explicit solution  to the hat relation~(\ref{N-HATrelation}) is given by 
\begin{eqnarray}
\label{eq:rep1}
& & a^{(N)}_{jk} =  
\left\{\begin{array}{ll}
A^{\ot(j-1)}\ot\del\ot\one^{\ot(k-j-1)}\ot\eps\ot\one^{\ot(N-k-1)}
 & (j<k<N), \\
A^{\ot(j-1)}\ot\del\ot\one^{\ot(N-j-1)} 
 & (j<k=N), \\
A^{\ot(j-1)}\ot\one^{\ot(N-j)} & (j=k), \\
\left(\sum_{i=1}^{k-1} y_i A^{\ot(i-1)}\ot B\ot \one^{\ot(k-i-1)} 
 + y_k A^{\ot(k-1)}\right)\ot\eps\ot\one^{\ot(N-k-1)}
 & (k<j-1 =N), \\
\sum_{i=1}^{N-1} y_i A^{\ot(i-1)}\ot B\ot \one^{\ot(N-i-1)} + y_N A^{\ot (N-1)}  
 & (j-1=k=N), \\
0 & $(otherwise)$
\end{array}\right. \\
& &  \widehat a^{(N)}_{jk} =  
\left\{\begin{array}{ll} 
  \one^{\ot (k-1)}\ot\eps\ot\one^{\ot(N-k-1)}   & (k<j-1=N)  \\
  \one^{\ot(N-1)}   & (k=j-1=N) \\
  0 & $(otherwise)$.
\end{array}\right.
\label{eq:rep2}
\end{eqnarray}
The difference from the homogeneous case
appears in $a^{(N)}_{jk}$ for $j=N+1$. 
Indeed we retrieve the solution to 
 the homogeneous case  \cite{AAMP1,EFM}
 by setting $y_j=1$.
We change the definitions of $\ba^{(1)}$ and $\widehat\ba^{(1)}$ as
$\ba^{(1)}=\left(\begin{array}{c} 1 \\ y_1    \end{array}\right),\ 
\widehat\ba^{(1)}=\left(\begin{array}{c} 0 \\ 1 \end{array}\right)$
for  compatibility  with the general forms (\ref{eq:rep1}), (\ref{eq:rep2}) .

By a direct calculation, one can show that 
equations (\ref{eq:rep1}), (\ref{eq:rep2}) give 
a representation to the algebra defined by 
the hat relation (\ref{eq:hat}) i.e.
\begin{eqnarray}
\label{eq:alg1}
-x_j a_{jk} a_{j'k'} - x_k(a_{jk'} a_{j'k}-a_{jk} a_{j'k'})
  =  \widehat a_{jk} a_{j'k'}  -  a_{jk} \widehat a_{j'k'} 
  &   (j<j' \wedge k<k'), \\
- x_k(a_{jk'} a_{j'k}-a_{jk} a_{j'k'})
  =  \widehat a_{jk} a_{j'k'}  -  a_{jk} \widehat a_{j'k'} 
  &   (j=j' \wedge k<k'), \\
x_{j'} a_{j'k} a_{jk'} - x_k(a_{jk'} a_{j'k}-a_{jk} a_{j'k'})
  =  \widehat a_{jk} a_{j'k'}  -  a_{jk} \widehat a_{j'k'} 
  &   (j>j' \wedge k<k'), \\
-x_j a_{jk} a_{j'k'} =  \widehat a_{jk} a_{j'k'}  -  a_{jk} \widehat a_{j'k'} 
  &   (j<j' \wedge k\ge k'), \\
 0  =  \widehat a_{jk} a_{j'k'}  -  a_{jk} \widehat a_{j'k'} 
  &   (j=j' \wedge k\ge k'), \\
x_{j'} a_{j'k} a_{jk'}  =  \widehat a_{jk} a_{j'k'}  -  a_{jk} \widehat a_{j'k'} 
  &    (j>j' \wedge k\ge k') .
\end{eqnarray}
 It is straightforward to prove that
the above relations are satisfied simply  by 
 substituting equations (\ref{eq:rep1}) and  (\ref{eq:rep2}).
In appendix, we prove the first identity (\ref{eq:alg1}) as an example. For  $N=3$ and 4, the result can be written explicitly as 
\begin{eqnarray}
& &  \ba^{(3)}  =
  \left(\begin{array}{ccc}
   \one\ot\one & \del\ot\eps&
   \del\ot\one \\
   0 &  A\ot\one &
   A\ot\del\\
   0 & 0 &
   A\ot A \\
  y_1\eps\ot\one &  y_1 B \ot\eps  &  y_1 B \ot\one   \\ 
  \                  & +y_2 A \ot\eps & +y_2 A \ot B   \\ 
  \                  & \                  &  +y_3 A \ot A 
  \end{array}\right)  , \quad 
 \widehat\ba^{(3)} = 
  \left(\begin{array}{ccc}
  0 & 0 & 0 \\
  0 & 0 & 0 \\
  0 & 0 & 0 \\
  \eps\ot\one & \one\ot\eps & \one\ot\one    
    \end{array}\right)   \, ,
\end{eqnarray}
   \begin{eqnarray}
& &  \ba^{(4)}  =
  \left(\begin{array}{cccc}
   \one\ot\one\ot\one & \del\ot\eps\ot\one &
   \del\ot\one\ot\eps &  \del\ot\one\ot\one  \\
   0 &  A\ot\one\ot\one &
   A\ot\del\ot\eps &  A\ot\del\ot\one \\
   0 & 0 &
   A\ot A\ot\one &
   A\ot A\ot\del \\
   0 & 0 & 0 &
   A\ot A\ot A \\
  y_1\eps\ot\one\ot\one &
  y_1 B \ot\eps\ot\one &
  y_1 B \ot\one\ot\eps &
  y_1 B \ot\one\ot\one  \\ 
  \ &
  +y_2 A \ot\eps\ot\one &
  +y_2 A \ot B \ot\eps &
  +y_2 A \ot B \ot\one \\ 
  \ & \  &
  +y_3 A \ot A \ot\eps &
  +y_3 A \ot A \ot B  \\ 
  \ & \  & \  &
  +y_4 A \ot A \ot A
  \end{array}\right)  , \\
& &   \widehat\ba^{(4)} = 
  \left(\begin{array}{cccc}
  0 & 0 & 0 & 0 \\
  0 & 0 & 0 & 0 \\
  0 & 0 & 0 & 0 \\
  0 & 0 & 0 & 0 \\
  \eps\ot\one\ot\one &
  \one\ot\eps\ot\one &
  \one\ot\one\ot\eps &
  \one\ot\one\ot\one  
    \end{array}\right)   \, .
\end{eqnarray}

We now generalize  the form (\ref{DecompX2}) or (\ref{FORMALDecompX2}).
The form (\ref{eq:psipsipsi}) can be written as
the matrix product form   (\ref{eq:MP}) with the matrices
\begin{eqnarray}
 X^{(N)}_{j} =\langle j | X^{(N)}  |1\rangle  ,\quad 
 X^{(N)} =  \ba^{(N)} \star \cdots \star \ba^{(1)} \,.
\end{eqnarray}
 thanks to the ``sector specificity''\footnote{
When $k_1\cdots k_L$ does not belong to the sector $m'$, 
the trace (\ref{eq:psi-element}) is always 0.
This specificity is
 because the numbers of $\del$'s and $\eps$'s are different 
in the matrix product, see \cite{AAMP1} for details.}.

Let us consider the element (\ref{eq:psi-element})
for configurations $j_1\cdots j_L$ and $k_1\cdots k_L$
of the sectors  $(m_1,\dots,m_{N+1})$ and  $(m_1,\dots, m_N+m_{N+1})$
with $j_i<j_{i+1}$.
Since $a_{jk}a_{j'j'}=0$ for $j<j'\le N$ and $k\neq k'$,
we need to set $j_i=k_i$ for $i\le L-m_{N+1}$ 
so that  (\ref{eq:psi-element}) is nonzero, and 
we have 
\begin{eqnarray}
  {\rm Tr} \left( a_{11}\cdots a_{11}  a_{22}\cdots a_{22}
  \cdots a_{NN}\cdots a_{NN}  a_{(N+1)N}\cdots a_{(N+1)N} \right)   
 = y_N^{m_{N+1}}  .
\end{eqnarray}
Therefore the stationary weight of the configuration 
$j_1\cdots j_L$
is $ \prod_{n=1}^{N} y_n^{m_{n+1}+\cdots + m_{N+1}} $.
As far as we treat basic sectors,
 the matrix product $a^{(N)}_{j_1k_1}\cdots a^{(N)}_{j_Lk_L}$
 contains $a^{(N)}_{NN}=A^{\ot (N-1)}$ or 0.
This implies that the matrix product 
 is transformed into a tensor product of the form 
  $\eps^c A \del^c =  |c \ketket  \brabra c | $
 multiplied by a monomial  of $y_n$'s
(by applying the relation (\ref{eq:DEHP}) and $A^2=A$),  
otherwise it is 0.
Since ${\rm Tr} \eps^c A \del^c = 1$,
  the trace of $a^{(N)}_{j_1k_1}\cdots a^{(N)}_{j_Lk_L}$
  is a monomial of $y_n$'s or 0.
Note that the trace is not always finite if we consider 
 a {\it non} basic sector.

\section{Graphical construction of the stationary state}
 \label{sec:FMALalgo}

 We have found a solution to the generalized matrix product Ansatz
  that constructs conjugation matrices of the inhomogeneous $N$-TASEP.
  In this section, we first review  the algorithm
 that A. Ayyer and S. Linusson devised \cite{AL} to calculate the stationary weights
 by defining an inhomogeneous extension 
of the  seminal algorithm by  Ferrari and Martin\cite{FM}.
Then we show that
 the solution to the matrix product Ansatz
  of the previous section is equivalent to   Ayyer and Linusson's construction.

\subsection{Ayyer and Linusson's algorithm}\label{sec:AL}

\begin{figure}
\begin{center}
 \includegraphics[width=0.7\columnwidth]{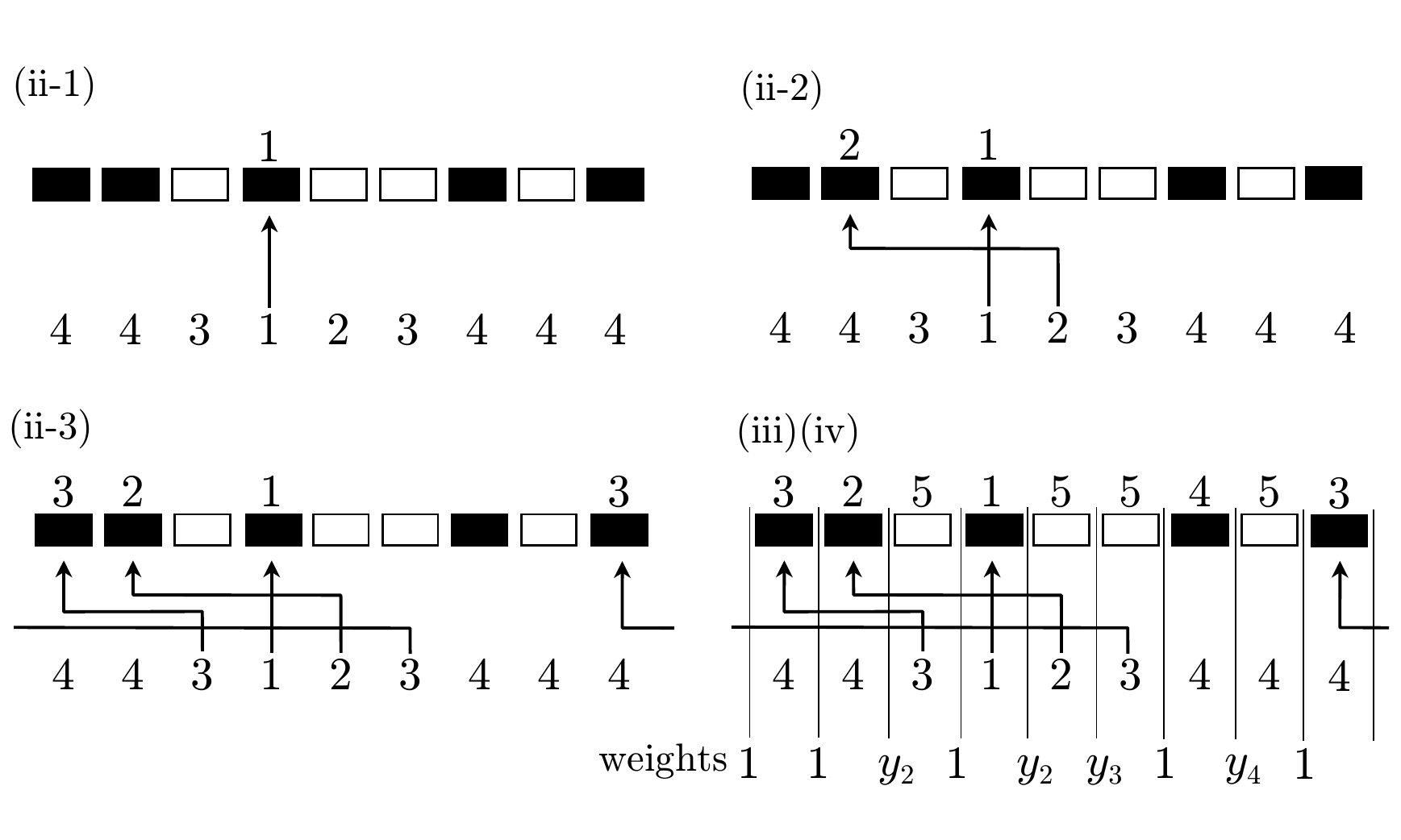}
\label{figAL}
\caption{An example of Ayyer and Linusson's algorithm.}
\end{center}
\end{figure}

 The   algorithm   of  Ayyer and Linusson \cite{AL}
  constructs the stationary state of the
  $N$-species sector $(m_1,\dots,m_{N+1})$
  from the $(N-1)$-species sector  $(m_1,\dots,m_N+m_{N+1})$.
It is provided by two   maps $F,W$
  from  an $(N-1)$-species configuration
 and a configuration consisting of black and white boxes,
 to an $N$-species configuration  and  a polynomial in $y_i$'s.
Figure \ref{figAL} is helpful to understand the algorithm.

\begin{itemize}
\item[(i)]
Let us set two lines.
 On the  upper line,
there are   $m_1+\cdots+m_N$ black boxes $\bsq$ 
 and $m_{N+1}$ white boxes $\sq$
 as $c_1\cdots c_L$ ($c_i=\blacksquare,\square$).
  On the  lower line, 
  we give a configuration $k_1\cdots k_L$ of the 
  $(N-1)$-species sector $(m_1,\dots,m_N+m_{N+1})$.
   Thus,  on the  lower line, 
  there are $m_\nu$  $\nu$'s ($1\le \nu \le N-1 $)
 and  $(m_N+m_{N+1})$  $N$'s. 
\item[(ii-1)]
Let $\{ i_1^{(1)}, \dots, i_{m_1}^{(1)}\}$
  be the positions of the $m_1$ 1's on the lower line. 
For the first 1 located at $i_1^{(1)}$, 
 find the nearest black box $c_{i'}=\blacksquare$ with $i'\le i_1^{(1)}$
   and put 1 on it.
If there is no such black box,
  put 1 on the rightmost black box.
For the second 1 located at $i_2^{(1)},$
find the nearest unoccupied black box $c_{i'}=\blacksquare$
 with $i'\le i_2^{(1)}$
   and put 1 on it.
If there is no such black box,
  put 1 on the rightmost unoccupied black box.
We draw an arrow from $i_1^{(1)}$ to the targeted black box.
(In the following procedure till (ii-$\nu$), we always 
draw an arrow in the same way, see figure (\ref{figAL}))
We iterate this procedure $m_1$ times, i.e.  
find the nearest unoccupied black box $c_{i'}=\blacksquare$
 with $i'\le i_\ell^{(1)}$
   for the $\ell$th 1   located at  $i_\ell^{(1)},$
   and put 1 on it
  or on the rightmost unoccupied black box
  if  $i'$ does not exist.
\item[(ii-2)]
 Let  $\{ i_1^{(2)}, \dots, i_{m_2 }^{(2)}\}$ 
   be the positions of the $m_2$ 2's on the lower line. 
There are $(m_2+\cdots + m_N)$
 unoccupied black boxes on the upper line.
 We  iterate the following procedure $m_2$ times:
find the nearest unoccupied black box $c_{i'}=\blacksquare$ 
with $i'\le i_\ell^{(2)}$
 for the $\ell$th  2  ($1 \le \ell \le m_2$),  and put  2 on it
  or on the rightmost unoccupied black box
  if  $i'$ does not exist.
\item[(ii-$\nu$)]
 In the same way, we go on for $\nu=3,4,\dots,(N-1)$.
 Let  $\{ i_1^{(\nu)}, \dots, i_{m_\nu }^{(\nu)}\}$ 
   be the positions of the $m_\nu$ $\nu$'s on the lower line. 
There are $(m_\nu+\cdots+m_N)$
 unoccupied black boxes remaining on the upper line.
We iterate the following procedure $m_\nu$ times: 
find the nearest unoccupied black box $c_{i'}=\blacksquare$
 with $i'\le i_\ell^{(\nu)}$
  for the $\ell$th  $\nu$ ($1 \le \ell \le m_\nu$), 
  and put $\nu$ on it
  or on the rightmost unoccupied black box
  if  $i'$ does not exist.
\item[(iii)] 
There are $m_N$ unoccupied black boxes remaining.
Put $N$s on them.
\item[(iv)]  
Put $(N+1)$s on the $m_{N+1}$ white boxes.
We have thus constructed a configuration 
  $F(c_1\cdots c_L,k_1\cdots k_L) $ 
   of the $N$-TASEP on the upper line, belonging to the sector $m$.
\item[(v)]
Define a vector from the stationary state
$ | \bar P_{m'}\rangle$   of 
the sector $m'$ as follows:
\begin{eqnarray}\label{eq:ALconjectured}
  | \bar P_m\rangle =\sum W(j_1\cdots j_L, k_1\cdots k_L)
   |  j_1\cdots j_L \rangle \langle  k_1\cdots k_L | \bar P_{m'}  \rangle .
\end{eqnarray}
The summation $\sum$ runs over
  $ j_1\cdots j_L $ and $k_1\cdots k_L$ 
 belonging to the sectors $m$ and  $m'$.
If there exits a configuration $c_1\cdots c_L$ such that 
\begin{eqnarray}\label{eq:Fck=j}
F(c_1\cdots c_L,k_1\cdots k_L)=j_1\cdots j_L,
\end{eqnarray}
the coefficient $W(j_1\cdots j_L, k_1\cdots k_L)$ is defined as 
the product of the following weights $\{w_1,\cdots, w_L\}$.
Draw a vertical line between each bond between site $i-1$ and $i$,
see figure \ref{figAL}.
When an arrow connecting two $\nu$'s on the upper and lower lines
in our figures, we say ``the value of the arrow is $\nu$'':
{\scriptsize $\!\!\!\!\!\!\!\!\!
\begin{array}{c} \nu \\ {\quad}^\uparrow\!\!\!\!-\!\!\!-\!\!\!{\ }_{|}\\
\quad\quad\  \nu
  \end{array}\!\!\!.$ }
Each weight $w_i$ is defined as
\begin{eqnarray}
w_i=
\left\{\begin{array}{ll}
  1 & (j_i\le N ), \\
  y_\ell & (j_i = N+1,\ \text{and $\ell$ is the minimal value of arrows
  that cross the $i$th vertical line}), \\
  y_N & (j_i = N+1,\ \text{and no arrow crosses the $i$th vertical line})  .
\end{array}\right.
\end{eqnarray}
Then we have $W(j_1\cdots j_L, k_1\cdots k_L)=\prod_{1\le i\le L}w_i$.
If there is no configuration  $c_1\cdots c_L$ such that equation 
(\ref{eq:Fck=j}) is satisfied,
we define $W(j_1\cdots j_L, k_1\cdots k_L)=0$.
\end{itemize}

Ayyer and Linusson conjectured that 
the form (\ref{eq:ALconjectured})
 gives the stationary state $|\bar P_m \rangle$ of the sector $m$ \cite{AL}.
There  the definition of the weight looks  different from our $W$,
 but  is  equivalent by multiplying it by a constant.

Figure \ref{figAL} gives an example of the algorithm
for the sectors $m=(1,1,2,1,4), m'=(1,1,2,5)$,
where the upper and lower lines are
\begin{eqnarray}
 c_1\cdots c_L  = \bsq\bsq\sq\bsq\sq\sq\bsq\sq\bsq ,\quad
  k_1\cdots k_L =  4 4 3 1 2 3 4 4 4 .
\end{eqnarray}
According to the algorithm, the configuration of the sector $m$
and the weight are obtained as 
\begin{eqnarray}
F ( \bsq\bsq\sq\bsq\sq\sq\bsq\sq\bsq,  4 4 3 1 2 3 4 4 4 ) 
= 3 2 5 1 5 5 4 5 3  ,\quad 
W (  3 2 5 1 5 5 4 5 3 ,  4 4 3 1 2 3 4 4 4 )
= y_2^2  y_3  y_4 .
\end{eqnarray}

\subsection{Equivalence of the matrix representation and 
 Ayyer and Linusson's algorithm}\label{sec:equivalence}

We  explain  the relation between 
 the matrix product representation  and 
 Ayyer and Linusson's algorithm
(following  \cite{AAMP1} for  the homogeneous case).
The elements   $a^{(N)}_{jk}$ act on
a basis vector  $|\mu_1,\dots,\mu_{N-1}\ketket
 =|\mu_1\ketket\ot\cdots\ot|\mu_{N-1}\ketket \in\A^{\ot (N-1)}$ as 
\begin{eqnarray}
\nonumber
& & a^{(N)}_{jk}  |\mu_1,\dots,\mu_{N-1} \ketket \\
\label{eq:local-action}
& & = \left\{\begin{array}{ll}
    |\mu_1 ,\dots,\mu_j-1 ,\dots,\mu_k+1 ,\dots ,
     \mu_{N-1} \ketket  &
       ( j<k<N,\mu_1=\cdots=\mu_{j-1}=0,\mu_j>0),   \\
    |\mu_1 ,\dots,\mu_j-1 , \dots , \mu_{N-1} \ketket  &
       ( j<k=N,\mu_1=\cdots=\mu_{j-1}=0,\mu_j>0),   \\
    |\mu_1 ,\dots,\mu_{N-1} \ketket  &
     (  j=k ,\mu_1=\cdots=\mu_{j-1}=0) , \\
   y_{{ \min \{\ell,k\} } }   |\mu_1 , \dots,  \mu_k+1,\dots, \mu_{N-1} \ketket
     & (k<j-1=N,\mu_1=\cdots=\mu_{\ell-1}=0,\mu_\ell>0) ,  \\
   y_\ell |\mu_1   ,\dots, \mu_{N-1} \ketket
   & (j-1=k=N,\mu_1=\cdots=\mu_{\ell-1}=0,\mu_\ell>0),  \\
   0 & \text{(otherwise)} .
\end{array}\right.
\end{eqnarray} 
Let  $j_1\cdots j_L$ and $k_1\cdots k_L$ belong to
basic sectors $m$ and $m'$.
If the vector $  | \mu_1,\dots,\mu_{N-1} \ketket $
is not killed by a  matrix product $a_{j_1k_1}\cdots a_{j_Lk_L}$,
the series of matrices $\{a_{j_1k_1},\dots, a_{j_Lk_L}\}$
 give a series (``trajectory'') as 
\begin{eqnarray}\label{eq:trajectory}
\nonumber
  | \mu_1,\dots,\mu_{N-1} \ketket 
  \stackrel{a^{(N)}_{j_Lk_L}}{\longmapsto}    
 v_L | \mu^L_1,\dots,\mu^L_{N-1} \ketket
   \stackrel{a^{(N)}_{j_{L-1}k_{L-1}}  }{\longmapsto}     
 v_{L-1}v_L | \mu^{L-1}_1,\dots,\mu^{L-1}_{N-1} \ketket \mapsto \\   \cdots \ \mapsto \ 
v_2\cdots v_L | \mu^2_1,\dots,\mu^2_{N-1} \ketket    
\ \stackrel{a^{(N)}_{j_1k_1}}{\longmapsto} \ 
v_1\cdots v_L | \mu^1_1,\dots,\mu^1_{N-1} \ketket ,
\end{eqnarray}
where we have set 
\begin{eqnarray}
a^{(N)}_{j_ik_i} | \mu^{i+1}_1,\dots,\mu^{i+1}_{N-1} \ketket 
 = v_i   | \mu^i_1,\dots,\mu^i_{N-1} \ketket  \quad
 ( \mu^{L+1}_\nu = \mu_\nu) .
\end{eqnarray}
Since $a^{(N)}_{jk}$ increases 
$\mu_k\mapsto \mu_k+1$ and decreases $\mu_j\mapsto \mu_j-1$
(see the action (\ref{eq:local-action})),
and we have $\#\{j_i=\nu\}=\#\{k_i=\nu\}$ for $\nu\le N-1$,  
we find $|\mu^1_1,\dots,\mu^1_{N-1}\ketket
=|\mu_1,\dots,\mu_{N-1}\ketket$.
In other words, $| \mu_1,\dots,\mu_{N-1}\ketket$
is an eigenvector of the matrix product
with a nonzero eigenvalue.
If a  trajectory  (\ref{eq:trajectory}) is given, 
one notices  that it is unique  and  
$a^{(N)}_{j_1k_1}\cdots a^{(N)}_{j_Lk_L}
|\mu'_1,\dots,\mu'_{N-1}\ketket =0$
for $(\mu'_1,\dots,\mu'_{N-1})\neq (\mu_1,\dots,\mu_{N-1} )$.
We regard $\mu_\nu$ in the vector $| \mu_1,\dots,\mu_{N-1}\ketket$
as the number of arrows with value $\nu$.
The trajectory gives one graph of arrows which is 
the same as obtained by Ayyer and Linusson's algorithm
since the local action (\ref{eq:local-action}) 
is compatible with it.
(For example, when  $j<k<N$, 
the action of $a^{(N)}_{jk}$ decreases $\#$ of arrows with value $j$
and increases $\#$ of arrows with value $k$.
This is allowed only if there is no arrow with value $\nu<j$,
otherwise it kills the vector.)
We can also see a compatibility of the local action 
 for the coefficient  $v_i=w_i$ as well as arrows,
 and thus the nonzero eigenvalue is identical to
  $W(j_1\cdots j_L,k_1\cdots k_L)$:
\begin{eqnarray}
   {\rm Tr} \left(a^{(N)}_{j_1k_1}\cdots a^{(N)}_{j_Lk_L}\right)
=    W(j_1\cdots j_L,k_1\cdots k_L)  . \label{eq:Tr=W}
\end{eqnarray}
When there is no nonzero eigenvalue, this equation 
is also true since $W(j_1\cdots j_L,k_1\cdots k_L)  =0$.

For example,  for configurations 
$  j_1 \cdots j_L = 3 2 5 1 5 5 4 5 3 , \, 
    k_1 \cdots k_L = 4 4 3 1 2 3 4 4 4  $
as in  figure \ref{figAL},
the matrix product $a_{j_1k_1}\cdots a_{j_Lk_L}$
gives a trajectory 
\begin{eqnarray}
\nonumber
                 | 0,0,1 \ketket \ \stackrel{a^{(4)}_{34}}{\longmapsto}\  
                 | 0,0,0 \ketket \ \stackrel{a^{(4)}_{54}}{\longmapsto}\  
        y_4    | 0,0,0 \ketket \ \stackrel{a^{(4)}_{44}}{\longmapsto}\  
                 | 0,0,0 \ketket \ \stackrel{a^{(4)}_{53}}{\longmapsto}\  
    y_3y_4   | 0,0,1 \ketket \ \stackrel{a^{(4)}_{52}}{\longmapsto}\  
y_2y_3y_4   | 0,1,1 \ketket \\ \ \stackrel{a^{(4)}_{11}}{\longmapsto}\  
y_2y_3y_4   | 0,1,1 \ketket \ \stackrel{a^{(4)}_{53}}{\longmapsto}\  
y^2_2y_3y_4| 0,1,2 \ketket \ \stackrel{a^{(4)}_{24}}{\longmapsto}\  
y^2_2y_3y_4| 0,0,2 \ketket \ \stackrel{a^{(4)}_{34}}{\longmapsto}\  
y^2_2y_3y_4| 0,0,1 \ketket ,
\end{eqnarray}
and we have  $a_{j_1k_1}\cdots a_{j_Lk_L}  | 0,0,1 \ketket
=y^2_2y_3y_4 | 0,0,1 \ketket$,
and we have ${\rm Tr}(a_{j_1k_1}\cdots a_{j_Lk_L})=y^2_2y_3y_4$.

\section{Concluding remarks}\label{sec:conc}

 In this work, we applyed the generalized  matrix product
 Ansatz to represent the stationary weights of  the inhomogeneous $N$-species
 TASEP.  We also explained  that  our solution to  the Ansatz is equivalent to 
 Ayyer and  Linusson's  combinatorial algorithm. 
 Our analysis was motivated by some conjectures proposed by 
 Lam and Williams \cite{LW}. For example, for $L=N+1$, they
 claim that the stationary probability is
  a polynomial with respect to
  the hoping rates  with non-negative integer coefficients
  and is a non-negative integral sum of Schubert polynomials.
  The first observation is an obvious consequence of 
  Kirchhoff's matrix tree formula \cite{ZS}.
  (Note that this is also an outcome
  of our  matrix  product solution 
  because all operators have positive entries.)
  However, the relation with 
  Schubert polynomials is still unclear. 
  Another interesting  problem would be to  extend 
  the present study to the partially asymmetric case and to classify
  the multi-species  systems with arbitrary  inhomogeneous hopping rates that
  can be solved by the matrix product Ansatz. Finally, we note
  here that the inhomogeneities are linked to the particles rather than to
  the underlying lattice. Extending our approach to models with lattice
  defects (such as the Janowsky and  Lebowitz model in which
  the insertion of a slow bond  can generate a shock  \cite{LebJan})
  remains  a very challenging open question.

\section*{Acknowledgments}
We thank Arvind Ayyer for introducing us to this problem
and for interesting discussions.
C Arita is a JSPS fellow for research abroad.

\appendix
\section{Proof of equation (\ref{eq:alg1})   } 

Here we show the first case (\ref{eq:alg1}) of  the algebra.

\begin{itemize}

\item[] $\!\!\!\!\!\!\!$  The case when \fbox{$j\neq k,j'\le N$}. We have 
\begin{eqnarray}
   a_{j\kappa} =0 \quad 
  \text{or\quad $j$th component of $a_{j\kappa}$ is } \delta
  \quad (\text{for\ } \kappa=k,k'), \\
   a_{j'\kappa} =0 \quad 
  \text{or\quad $j$th component   of $a_{j'\kappa}$ is } A
  \quad (\text{for\ } \kappa=k,k') .
\end{eqnarray}
In any cases, we find $a_{jk}a_{j'k'} = a_{j'k}a_{jk'} = 0$,
and thus the left-hand side is 0.
The right-hand side is also 0
 thanks to $\widehat a_{jk}=\widehat a_{j'k'}=0$.

\item[] $\!\!\!\!\!\!\!$ The case when   \fbox{$j=k,j'\le N$}. 
 The left hand side is $-x_j a_{jk'}a_{j'j}=0$ 
thanks to $a_{j'j}=0$.
The right-hand side is again 0
 thanks to $\widehat a_{jk}=\widehat a_{j'k'}=0$.

\item[] $\!\!\!\!\!\!\!$ The case when  \fbox{$j<k,j'=N+1$}.  Since 
\begin{eqnarray}
  a_{jk} a_{j'k'} = a_{jk'} a_{j'k} = 
   y_j A^{\ot(j-1)}\ot\del\ot\one^{\ot(k-j-1)}
  \ot\eps\ot\one^{\ot(k'-k-1)}\ot\eps\ot\one^{\ot(N-k'-1)}, 
\end{eqnarray}
the left hand side is calculated as 
\begin{eqnarray}
   - A^{\ot(j-1)}\ot\del\ot\one^{\ot(k-j-1)}
  \ot\eps\ot\one^{\ot(k'-k-1)}\ot\eps\ot\one^{\ot(N-k'-1)} ,
\end{eqnarray}
which agrees with the right-hand side.
(We read $\cdots\one^{\ot(N-k-1)}\ot\eps\ot\one^{\ot(-1)}
=\cdots\one^{\ot(N-k-1)}$ for $k'=N$.)

\item[] $\!\!\!\!\!\!\!$ The case when  \fbox{$j=k,j'=N+1$}. 
The left-hand side is calculated as
\begin{eqnarray}
    - x_j a_{jk'} a_{j'j}   =
 - A^{\ot(j-1)}\ot\one^{\ot(k'-j)}\eps\ot\one^{\ot(N-k'-1)} ,
\end{eqnarray}
which agrees with the right-hand side.

\item[] $\!\!\!\!\!\!\!$ The case when  \fbox{$j>k,j'=N+1$}. 
We have $a_{jk}a_{j'k'}=0$ thanks to $a_{jk}=0$.
Since the $k$th   component  of each term of $a_{j'k}$ is $\eps$, and 
$a_{jk'}$=0 or the $k$th  component 
$a_{jk'}$ is $A$, we also have $a_{jk'}a_{j'k}=0$.
Thus the left hand side is 0.
The right hand side is also 0 thanks to
 $a_{jk}=\widehat a_{jk} =0$.

\end{itemize}

\end{document}